\documentclass[aps,amsmath]{revtex4}

\newcommand{\ce}[1]{{\color{black} #1}}
\newcommand\R{\zeta}	
	
	\usepackage{amsmath}
	\usepackage{amsfonts}
  	\usepackage{amssymb}
	\usepackage{makeidx}
	\usepackage{amsfonts}
	\usepackage[ansinew]{}
	\usepackage[usenames,dvipsnames]{pstricks}
	\usepackage{epsfig}

\usepackage{xcolor}

\renewcommand\theequation{\arabic{equation}}

\newcommand{\be}{\begin{equation}}
\newcommand{\ee}{\end{equation}}
\newcommand{\bea}{\begin{eqnarray}}
\newcommand{\eea}{\end{eqnarray}}

\newcommand\SPD{\mathrel{\stackrel{\makebox[0pt]{\mbox{\normalfont\tiny (3)}}}{\Delta}}}

\makeindex

\begin{document}


\title{\ce{Model-independent approach to effective sound speed in multi-field inflation}}

\author{Antonio Enea Romano${}^{1,2}$}
\author{Krzysztof Turzy\'nski${}^3$}
\author{Sergio Andr\'es Vallejo-Pe\~na${}^{1,2}$}

\affiliation{
${}^{1}${ICRANet, Piazza della Repubblica 10, I--65122 Pescara} \\
${}^{2}$Instituto de Fisica, Universidad de Antioquia, A.A.1226, Medellin, Colombia \\
${}^3$ Institute of Theoretical Physics, Faculty of Physics, University of Warsaw, Pasteura 5, 02-093 Warsaw,
Poland\\
}

\begin{abstract}
 For any physical system satisfying the Einstein's equations, the comoving curvature perturbations satisfy an equation involving 
 the momentum-dependent effective sound speed, 
 valid for any system with a well defined energy-stress tensor, 
 including multi-fields models of inflation.  
 \ce{
 We derive a general model-independent formula for the effective sound speed
 of comoving adiabatic perturbations, valid for a generic  
 field-space metric, without assuming any approximation to integrate 
 out entropy perturbations, 
 but
 expressing the momentum-dependent effective sound speed in terms of the components of the total energy-stress tensor. 
%
%
 As an application,
 we study a number of two-field models with a kinetic coupling between
 the fields, identifying 
 the single curvature mode of the effective theory
 and showing
 that momentum-dependent effective sound speed fully accounts for the predictions for the power spectrum
 of curvature perturbations.
 Our results show that the momentum-dependent effective sound speed is a convenient scheme for describing all inflationary models that admit a single-field
 effective theory, including the effects of entropy perturbations present in multi-fields systems.
 }
\end{abstract}


\maketitle
\section{Introduction}
The study of cosmological perturbations is one of the foundations of modern cosmology, since it allows to make quantitative predictions for different observables such as the 
characteristics of the 
cosmic microwave background radiation or  large scale structure formation. In the simplest models of inflation, with a single scalar field minimally coupled to gravity, the scalar field is driving the accelerated expansion of the Universe, and its perturbations induce metric perturbations which, in the comoving gauge, obey an evolution equation containing a Laplacian whose coefficient is called sound speed.
In these models, the sound speed is only a function of time, but it has been shown \cite{Romano:2018frb} that a similar equation, but with a space- or momentum-dependent sound speed, is satisfied by an adiabatic perturbation in an arbitrary physical system satisfying Einstein's equations, including multi-field models and modified gravity models.


In general, a given mode of adiabatic perturbations can receive contributions
from different degrees of freedom coupled to that mode. 
However, there exist a broad class of models,
including models with a strong kinetic coupling
between the adiabatic and entropy perturbations,
in which the mode of adiabatic perturbations responsible
for generation of observable CMB anisotropies evolves
independently of other modes. There has been
an extensive effort to identify situtations in which
complex models of inflation can be effectively
derscribed with a single-field effective theory 
with possible corrections \cite{Achucarro:2010da, Achucarro:2011yc, Shiu:2011qw, Avgoustidis:2011em, Achucarro:2012sm, Cespedes:2012hu, Avgoustidis:2012yc, Chen:2012ge, Pi:2012gf, Gao:2012uq, Achucarro:2012yr, Collins:2012nq, Burgess:2012dz, Gwyn:2012mw, Noumi:2012vr, Dimastrogiovanni:2012st, Bartolo:2013exa, Garcia-Saenz:2018vqf, Durakovic:2019kqq, Pinol:2020kvw}.


In this paper, we show that the evolution of that effective
adiabatic mode is correctly described within the formalism
of momentum-dependent effective sound speed,
discussing the notion of effective single-field theory for
inflationary perturbations and providing a set of numerical
calculations corresponding to specific two-field inflationary
models that have attracted considerable attention.

The paper is organized as follows. In Section \ref{sec:mess},
we briefly introduce the formalism of momentum-dependent sound speed. 
In Section \ref{sec:eff},
we analyze decoupling of heavy degrees of freedom and
calculate the sound speed in models with a constant turning rate 
of the inflationary trajectory from the geodesic line.
In Section \ref{sec:lf1}, we discuss the normalization of
perturbations and appropriate initial conditions in single-field effective theories
by means of
the Liouville formula.
Section \ref{sec:num} is devoted to numerical examples
corroborating our analytical calculations. After a short discussion
of the results in Section \ref{sec:diss}, we conclude
in Section \ref{sec:con}. 
Appendices contain more technical aspects of our derivations:
a calculation
of the effective sound speed in two-field models with
arbitrary field-space metric, as well as the generalization of
the Liouville formula to multi-field models and the resulting
discussion of the initial conditions for the perturbations.

\section{Momentum effective sound speed}
\label{sec:mess}
\subsection{Effective equation of motion}

It has recently been shown \cite{Romano:2018frb} 
that for any system satisfying Einstein's equations
the evolution of the adiabatic perturbation $\R$ can be described by means of a single differential equation
\begin{align}
\ddot{\R} + \frac{\partial_t(Z^2)}{Z^2} \dot{\R} &- \frac{v_s^2}{a^2} \SPD \R + \frac{v_s^2}{\epsilon}\SPD \Pi +  \frac{1}{3 Z^2}\partial_{t}\left( \frac{Z^2}{H \epsilon} \SPD \Pi \right)= 0 \, . \label{RcttPi}
\end{align}
where $Z^2\equiv\epsilon a^3/v_s^2$ and an effective space-dependent sound speed (SESS) has been defined as
\begin{equation}
v_s^2 (t,x^i) \equiv \frac{\delta P_c(t,x^i)}{\delta\rho_c(t,x^i)} \, ,\label{vs}
\end{equation}
where $\delta\rho_c$ and $\delta P_c$ are the energy density and pressure
perturbations in the comoving gauge, respectively.

In this picture, the entropy perturbations do not appear explicitly in the equation for adiabatic perturbations, and are `hidden' in the SESS. This can
be understood by comparing (\ref{vs}) with the result of the
standard approach \cite{Kodama:1985bj}, in which entropy perturbations $\Gamma$ are defined by
\begin{align}
\delta P_c(t,x^i) &= c_s(t)^2 \delta\rho_c(t,x^i) + \Gamma(t,x^i) \, , \label{entropy} \end{align}
where $c_s$ is interpreted as sound speed and is a function of time only.
Combining eqs.~(\ref{entropy}) and (\ref{vs})
we get the relation between SESS and entropy perturbations:
\begin{align}
v_s^2 &= c_s^2\left[ 1 + \frac{\Gamma}{2 H \epsilon \left(\dot{\R} + \frac{1}{3H\epsilon} \SPD \Pi\right)}\right]^{-1} \, . \label{vcgamma}
\end{align}
In the momentum space, one can similarly write a single differential equation
for the Fourier components of the adiabatic perturbations:
\begin{align}
\ddot{\R}_k + \left(3H+\frac{\partial_t(\tilde{Z}_k^2)}{\tilde{Z}_k^2}\right) \dot{\R}_k & + \frac{\tilde{v}_k^2}{a^2} k^2 \R_k - \frac{\tilde{v}_k^2}{\epsilon}k^2 \Pi_k -  \frac{1}{3 \tilde{Z}_k^2}\partial_{t}\left( \frac{\tilde{Z}_k^2}{H \epsilon} k^2 \Pi_k \right)= 0  \, , \label{Rckeq}  
\end{align}
where the momentum-dependent effective sound speed (MESS) now reads:
\begin{equation}
\tilde{v}_k^2(t) \equiv \frac{\delta P_{c,k}(t)}{\delta\rho_{c,k}(t)} \, , \label{ck}
\end{equation}
and $\delta \rho_{c,k}(t)$ and $\delta P_{c,k}(t)$ are Fourier components
of the energy density perturbations and pressure in the comoving gauge, respectively, and
$ \tilde{Z}_k^2\equiv\epsilon /\tilde{v}_k^2$.
In this paper we will consider scalar fields with isotropic EST,
for which eq.~(\ref{Rckeq}) simplifies to
\begin{align}
\ddot{\R}_k + \left(3H+\frac{\partial_t(\tilde{Z}_k^2)}{\tilde{Z}_k^2}\right) \dot{\R}_k & + \frac{\tilde{v}_k^2}{a^2} k^2 \R_k = 0  \, . \label{Rckeq2}  
\end{align}
It can be shown that eq.~(\ref{Rckeq2})
reduces to the Sasaki-Mukhanov equation when $\tilde{v}_k$ is a function of \textit{time only}.
It is important to note that the MESS $\tilde{v}_k(t)$  defined in eq.(\ref{ck}) is not simply the Fourier transform of the SESS $v_s(x^{\mu})$ defined in eq.~(\ref{vs}), because the product of the  Fourier transforms of two functions is the transform of the convolution of the two functions.




\subsection{Solution of the effective equation of motion}
\label{sec:sol}

In order to solve eq.~(\ref{Rckeq2}), one has to know the
time evolution of $\tilde{v}_k^2$. With a simple
phenomenological assumption that this quantity evolves 
as a power law of the scale factor, i.e.~$\tilde{v}_k^2=V_0^2 a^p$, we can solve eq.~(\ref{Rckeq2}) in the limit $|\dot{H}|\ll H^2$, obtaining:
\begin{align}
\R = D_{i} \mathcal{A}^{\frac{1}{2}(p-3)} H^{(i)}_{\left| \frac{p-3}{p-2}\right|}\!\left( \frac{\mathcal{A}^{\frac{1}{2}(p-2)}}{\frac{1}{2}|p-2|}\right)
\,,\qquad i=1,2\, ,
\label{eq:Rsol}
\end{align}
where $H^{(i)}_\mu$ are Hankel functions of the first and second kind, respectively.
$a=\kappa \mathcal{A}$ with $\kappa^{2-p}=V_0^2k^2/H^2$ and $V_0^2>0$; for $V_0^2<0$ the argument of the Hankel function has to be multiplied by the imaginary unit. In the special case $p=2$, the solution is
\begin{align}
\R = D_\pm \mathcal{A}^{-\frac{1}{2}\pm\frac{1}{2}\sqrt{1-\frac{4V_0^2k^2}{H^2}}} \, .    
\end{align}
The late-time asymptotic behavior of the solution (\ref{eq:Rsol}) depends on the value of $p$. 

For $p<2$, the
argument of the Hankel function goes to zero as $a$ increases to infinity and using $H_\mu^{(1)}(\xi)\sim -\mathrm{i}(\Gamma(\mu)/\pi)(2/\xi)^\mu$ for small $|\xi|$, where $\Gamma$ is the Euler gamma function, we obtain for the positive frequency solution
\begin{align}
\R \sim -\frac{\mathrm{i} D_1 (2-p)^{\frac{3-p}{2-p}} \Gamma\left( \frac{3-p}{2-p}\right)}{\pi} = \mathrm{const} \, .
\end{align}
Thus, for $p<2$ there is a freezing mode of the curvature perturbation, irrespective of the sign of $V_0^2$. 

For $p>2$, the
argument of the Hankel function goes to infinity as $a$ increases to infinity and the asymptotic behavior of eq.~($\ref{Rckeq2}$) becomes
\begin{align}
    \R \sim \mathcal{A}^{\frac{p}{4}-1} \mathrm{exp}\!\left(\frac{\mathrm{i}\mathcal{A}^{\frac{1}{2}(p-2)}}{\frac{1}{2}(p-2)}\right) \, 
\end{align}
With $V_0^2>0$, we obtain decaying solutions for $2<p<4$ and growing solutions for $p>4$; all solutions oscillate. For $V_0^2<0$, there is an exponential growth of the solution.
These cases do not admit a freezing solution for the
curvature perturbation.

\begin{figure}[t]
    \centering
    \includegraphics[width=.44\textwidth]{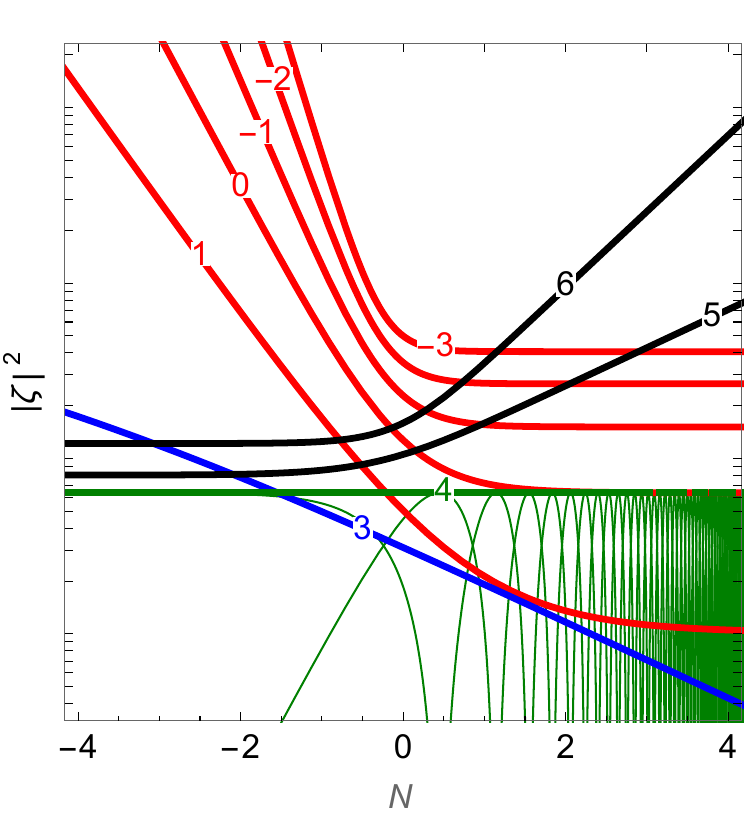}
    \caption{Evolutions of the amplitude of curvature perturbation $\R$ given in eq.~(\ref{eq:Rsol}) for different sound speeds $\tilde{v}_k^2\propto a^p$ with different values of $p$ is shown as thick lines. Color coding corresponds to late-time behavior: freezing (red), decaying (blue) and growing (black); the special case of $p=4$ is shown in green. Thin green lines indicate the real and imaginary part of of $\R$ for $p=4$. Normalization of $\R$ is arbitrary.}
    \label{fig:time_ev}
\end{figure}

In Figure~\ref{fig:time_ev}, we show the time evolution of
curvature perturbations given by eq.~(\ref{eq:Rsol}).

\section{Effective equation of motion {\em vs} full theory in multi-field models}
\label{sec:eff}

As a particular example, we will consider models involving $N=2$ scalar fields minimally coupled to Einstein gravity, and whose action reads:
\be
S=\int {\rm d}^4 x \sqrt{-g} \left[-\frac12 G_{IJ}\left(\phi^K\right) \partial_{\mu} \phi^I \partial^{\mu} \phi^J-V\left(\phi^K\right) \right]\,.
\label{action}
\ee
In eq.~(\ref{action}), uppercase Latin letters refer to the field space directions and summation over repeated indices is assumed.
 It is 
 convenient to project the evolution of homogeneous fields and the perturbations in the field space onto the adiabatic/entropic basis $(e_{\sigma}^I,e_{s}^I)$ \cite{Gordon:2000hv,GrootNibbelink:2001qt}, where  $e_{\sigma}^I \equiv \dot \phi^I /{\dot \sigma}$ is the unit vector pointing along the background trajectory in field space, and where $e_s^I$ is such that the basis $(e_{\sigma}^I,e_{s}^I)$ is orthonormal and right-handed for definiteness; 
 the velocity of the system in the field space reads $\dot\sigma=(G_{IJ}\dot\phi^I\dot\phi^J)^{1/2}$.
 The adiabatic perturbation $Q_{\sigma} \equiv e_{\sigma I} Q^I$ is directly proportional to the comoving curvature perturbation $\zeta=\frac{H}{{\dot \sigma}} Q_{\sigma}$, 
while the genuine multifield effects are embodied by the entropic fluctuation $Q_s$, perpendicular to the background trajectory.

In this basis, the equations of motion take the form
\begin{eqnarray}
\label{Qsigma}
 \ddot{Q}_{\sigma}+3H
 \dot{Q}_{\sigma}+\left(\frac{k^2}{a^2}+m_{\sigma}^2\right)  Q_{\sigma} &=& {2 H \eta_\perp \dot{Q}_{s}}
-\left(\frac{\dot H}{H}
+\frac{V_{,\sigma}}{\dot \sigma }\right)  2 H \eta_\perp\, Q_{s}\,, \\
\ddot{Q}_s+3H \dot{Q}_{s}+\left(\frac{k^2}{a^2}+m_{s}^2\right)  Q_{s}&=&-2 \dot \sigma \eta_\perp \dot{\zeta}\,,
\label{Qs}
\end{eqnarray}
where
\begin{equation} \label{etaperpdefinition}
\eta_\perp  \equiv -\frac{V_{,s} }{H \dot \sigma}
\end{equation}
is the dimensionless parameter, describing the rate (in Hubble times) at which the trajectory in the field space deviates from a geodesic line \cite{GrootNibbelink:2001qt}. Here $V_{,s}  \equiv e_{s}^I V_{,I}$, the adiabatic mass (squared) is given by $m_{\sigma}^2/H^2=-\frac{3}{2}\epsilon_2+\ldots$
with the slow-roll parameters given by $\epsilon_1 \equiv -\frac{\dot H}{H^2}$, $\epsilon_2 =\frac{\dot \epsilon_1}{H \epsilon_1}$ and the dots representing terms of higher order in the slow-roll parameters, and the entropic mass squared reads $m_s^2=V_{;ss}-2( H \eta_\perp)^2$.

In order to connect the system of equations of motion (\ref{Qsigma})
and (\ref{Qs}) to the effective equation (\ref{Rckeq}),
we note that
\begin{eqnarray}
\label{eq:alpha2f}
\delta\rho_{c,k}(t) &=& -\frac{\dot{\R}}{H}\dot\sigma = - \frac{H^2\dot{\sigma}^2}{\dot{H}} \frac{k^2}{a^2H^2} \Psi - 2\eta_\perp H Q_s \\
\label{eq:beta2f}
\delta P_{c,k}(t) &=& \delta\rho_{c,k}(t) + 2\eta_\perp H Q_s = - \frac{H^2\dot{\sigma}^2}{\dot{H}} \frac{k^2}{a^2H^2} \Psi \, ,
\end{eqnarray}
where $\Psi$ is the Bardeen potential.
Inserting (\ref{eq:alpha2f}) and (\ref{eq:beta2f}) into (\ref{ck}),
we find that 
\begin{equation}
\tilde{v}_k^{-2} = 1-\frac{2\eta_\perp H^2 Q_s}{\dot{\R} \dot{\sigma}} \, \label{vk2} . 
\end{equation}
Plugging (\ref{vk2}) into (\ref{RcttPi}), we find that the latter equation, upon setting $\Pi=0$, which is appropriate for the system of scalar fields, is equivalent to (\ref{Qsigma}), {\em i.e.} it describes the
evolution of the adiabatic perturbations if it is supplemented by (\ref{Qs}) that dictates the evolution of the entropy perturbations.


\subsection{Momentum-dependent sound speed and effective field theory of inflation}

An important comment is now in order.
For any classical solution of the equations of motion for
the perturbations~(\ref{Qsigma}) and~(\ref{Qs}),
it is always possible to determine $\tilde{v}_k^2$ from (\ref{vk2})
and then the adiabatic perturbation $\R$ satisfies the effective equation
of motion~(\ref{Rckeq2}). Different choices of initial conditions for
a multi-field system would lead to different functions~$\tilde{v}_k^2$.
In order to account for all degrees of freedom in a  multi-field system,
it appears necessary to define as many momentum-dependent effective sound 
speeds as the number of the fields.

However, a considerable simplification arises if the amplitudes of the
perturbations other than the final adiabatic perturbation decay significantly
on super-Hubble scales. For concreteness, let us discuss this point for
two-field models.

Equations of motion~(\ref{Qsigma}) and~(\ref{Qs}) have to
be supplemented with initial conditions for the fields.
(One often adopts initial conditions with vanishing either entropic
or adiabatic perturbations, but we argue in Appendix~\ref{app:bdini}
that other choices may be more natural.) 
As a result, one obtains two solutions, $\R_1$ and~$\R_2$,
corresponding to different, orthogonal initial conditions.
Because they correspond to different quantum degrees of freedom,
for calculation of the power spectrum they should be added
in quadratures, $|\R|^2=|\R_1|^2+|\R_2|^2$. However, if the late-time
super-Hubble behavior of the modes is dominated by a single degree
of freedom, one can perform a unitary transformation~$U$, such that:
\be
\left( \begin{array}{c} \tilde{\R}_1 \\ \tilde{\R}_2 \end{array} \right) = U\left( \begin{array}{c} \R_1 \\ \R_2 \end{array} \right)
\ee
with $\tilde{\R}_2\to0$ at late times; then $|\R|^2=|\tilde{\R}_1|^2$
fully accounts for the adiabatic power spectrum. An identical transformation
can then be performed for entropy modes. In such a case, we define
the momentum-dependent effective sound speed as one obtained
for $\tilde{\R}_1$ and the associated entropy perturbation.

While applying the procedure described above guarantees reproducing
the full time evolution of $\tilde{\R}_1$, specifying $\tilde{v}_k^2$
is not equivalent to formulating an effective single-field theory
of perturbations. This is because the matrix $U$ is defined
in terms of the late-time behavior of adiabatic perturbations
and this does not ensure a proper single-field normalization of
perturbations at early times, in sub-Hubble regime. It can readily
be seen for initial conditions $\tilde{\R}_{2,\mathrm{i}}=0$
and $|U_{11}|<1$. 

We can conclude that the usefulness of introducing MESS
consists in the possibility to account fully for time dependence
of the adiabatic perturbations,
even in cases in which a single-field effective theory
does not exist
On the other hand, if it does, then introducing MESS is equivalent
to formulating the effective theory to calculate the power
spectrum of the adiabatic perturbations.

There are several examples discussed in the literature, which admit
an effective single-field description and for which the predictions
for the power spectrum of adiabatic perturbation was calculated.
These examples are obtained in a two-field inflationary model, in which,
to make discussion easier, 
the inflationary trajectory exhibits a constant turning rate 
in the field space. 
Depending on that rate and on the mass parameters of the fields, several interesting cases in which the evolution of the perturbations differs significantly from the single-field scenario have been discussed over last decade. Later in section~\ref{sec:num}, we shall demonstrate
the usefulness of MESS beyond those examples.

\subsection{Examples }

\subsubsection{Geodesic trajectory}
If the trajectory in the field space follows a geodesic line, the entropy perturbations do not affect the adiabatic perturbations, which evolve as if the entropy perturbations were entirely absent. We can, therefore, set $Q_s=0$ in eq.~(\ref{vk2}) and conclude that the speed of adiabatic perturbations is that of light, $\tilde{v}^2_k=1$.

\subsubsection{Sourcing on super-Hubble scales}
\label{sec:sourcing}
If the amplitude of the entropy modes are not significantly smaller than those of after the adiabatic ones after Hubble-radius crossing and the trajectory in the field space does not follow a geodesic line, adiadiabatic perturbations are sourced by the entropy ones. The rate of this sourcing can be read from eq.~(\ref{eq:alpha2f}); as the first term on the r.h.s.\ is negligible on super-Hubble scales, we arrive at $\dot{\R}\dot{\sigma}\approx 2\eta_\perp H^2Q_s^2$ and the two terms in eq.~(\ref{vk2}) practically cancel. This can be interpreted as infinite sound speed. This should not come as a surprise, because on super-Hubble scales, the amplitude of the adiabatic perturbations grows coherently over distances exceeding the size of the horizon.

\subsubsection{Strongly coupled perturbations and sub-Hubble freeze-in}
\label{sec:strongly}

If the turn rate is large, $\eta_\perp\gg 1$ and slowly varying, 
the adiabatic and entropy 
perturbations exhibit interesting dynamics, leading to the adiabatic 
perturbations freezing in before the Hubble radius crossing and to 
enhancement of the power spectrum compared to the predictions of a 
single-field scenario with the same Hubble and slow-roll parameters
\cite{Tolley:2009fg,Cremonini:2010ua,Baumann:2011su,Garcia-Saenz:2018ifx,Fumagalli:2019noh}.
This happens after the amplitude of the more massive of the solutions of the system of eqs.~(\ref{Qsigma}) and (\ref{Qs}) becomes negligible and the lighter and more slowly changing mode becomes dominant. The relation between the adiabatic and entropy component of that mode can be read from (\ref{Qs}):
\begin{equation}
    \left(\frac{k^2}{a^2}+m_{s}^2\right)  Q_{s}=-2 \dot \sigma \eta_\perp \dot{\zeta} \, .
    \label{Qssim}
\end{equation}
Substituting eq.~(\ref{Qssim}) to (\ref{vk2}), we obtain:
\begin{equation}
    \tilde{v}_k^{-2} = 1+\frac{4\eta_\perp^2}{\frac{k^2}{a^2H^2}+\frac{m_s^2}{H^2}} \, .
    \label{eq:vsgel}
\end{equation}
If the sound speed of perturbations deviates significantly from one, the second term in eq.~(\ref{eq:vsgel}) must dominate; depending on the relative size of the two terms in the denominator, we arrive at:
\begin{equation}
    \tilde{v}_k^2 \approx \frac{m_s^2}{4\eta_\perp^2 H^2} \qquad\textrm{for}\,\,k/a\ll m_s
    \label{eq:vsgel2a}
\end{equation}
or
\begin{equation}
    \tilde{v}_k^2 \approx
    \frac{k^2}{4\eta_\perp^2 a^2H^2} \approx \frac{k^2\eta^2}{4\eta_\perp^2} \qquad\textrm{for}\,\, k/a\gg m_s \, .
\label{eq:vsgel2b}
\end{equation}
The first limit shown in eq.~(\ref{eq:vsgel2a}) corresponds to
{\em constant} reduced sound speed and has been extensively studied in the literature. The positive and negative frequency solutions of eq.~(\ref{Rckeq}) read:
\begin{equation}
\label{eq:ord0}
    \R = A_{\pm}e^{\mp \mathrm{i} \tilde{v}_k k\eta} \left( 1\mp\frac{\mathrm{i}}{\tilde{v}_k k\eta}\right)\,,
\end{equation}
where $A$ is a normalization constant and the symbol $\pm$ refers
to positive- and negative-frequency solutions.

The second limit shown in eq.~(\ref{eq:vsgel2b}) was first studied in \cite{Cremonini:2010ua} and later in \cite{Baumann:2011su}; 
because of the explicit dependence of $\tilde{v}_k$ on $k$, 
we shall refer to these models as models with modified dispersion relations. They correspond to our solution
(\ref{eq:Rsol}) with $p=-2$ and $V_0^2=k^2/4\eta_\perp^2H^2$.


The examples discussed in this subsection offer a route to 
a consistent interpretation of eq.~(\ref{vk2}) in a class
of multi-field models that allow an effective field theory with just one
field. If the amplitudes of all the perturbations except for the freezing-in 
adiabatic perturbations decay quickly, either because they are massive or, according to eq.~(\ref{Qssim}), the
entropy perturbations are suppressed after freeze-in of curvature perturbations, we can describe the evolution of the adiabatic
perturbations in the single-field model with an effective sound
spped $v_k$, which depends both on time and the wavenumber of the mode. 

In Section \ref{sec:num}, we shall present a set of numerical examples,
corroborating the assertion above and show that the predictions of the effective theory are consistent with those of the full theory for all times. But before we start comparing the full and the effective theory, we shall need a tool to translate the evoultion of the effective sound speed to the normalization of the power spectrum. This tool will be provided by the Liouville formula described in the following Section.

\section{Liouville formula}
\label{sec:lf1}

The Liouville formula states that for a function $y(\eta)$, which solves
the equation:
\begin{equation}
\label{eq:diff1}
\frac{\mathrm{d}^2u}{\mathrm{d}\eta^2}+b_1(\eta)\frac{\mathrm{d}u}{\mathrm{d}\eta}+b_0(\eta) u = 0 \,,
\end{equation}
where $b_1$ and $b_0$ are real-values functions, the Wronskian defined as:
\begin{equation}
\label{eq:wr0}
    W(\eta) \equiv u^\ast \frac{\mathrm{d}u}{\mathrm{d}\eta} - \left( \frac{\mathrm{d}u}{\mathrm{d}\eta}\right)^\ast u 
\end{equation}
satisfies:
\begin{equation}
\label{eq:wr}
W(\eta) = W(\eta_0)\, \mathrm{exp}\!\left( - \int_{\eta_0}^\eta b_1(\eta')\,\mathrm{d}\eta' \right) \, .
\end{equation}
In order to apply eq.~(\ref{eq:wr}) to (\ref{Rckeq2}),
we substitute $u= a\R$
and take the independent variable to be conformal time.
Eq.~(\ref{Rckeq2}) becomes:
\begin{equation}
    u'' + \left(\frac{\mathrm{d}}{\mathrm{d}\eta}\log \tilde{Z}_k^2\right) \left( u'+\frac{1}{\eta}u\right)+ \left(\tilde{v}_k^2 k^2-\frac{2}{\eta^2} \right) u = 0 \,,
    \label{Rckeq3}
\end{equation}
where we used de Sitter approximation $a\approx-1/H\eta$ with constant~$H$.
We obtain
\begin{equation}
\label{eq:wr2}
W(\eta) = W(\eta_0)\, \mathrm{exp}\!\left( - \int_{\eta_0}^\eta \left(\frac{\mathrm{d}}{\mathrm{d}\eta'}\log \tilde{Z}_k^2\right)\,\mathrm{d}\eta' \right) = W(\eta_0) \frac{\tilde{Z}_k^2(\eta_0)}{\tilde{Z}_k^2(\eta)}\, .
\end{equation}
Remembering that $\tilde{Z}_k^2=\epsilon/\tilde{v}_k^2$
and assuming that the slow-roll parameter~$\epsilon$ does not
change significantly in the time interval
between the time when the observed adiabatic modes are deep inside
the Hubble radius and the time of freeze-in, we obtain:
\begin{equation}
\label{eq:wr3}
    W(\eta)=W(\eta_0) \frac{\tilde{v}_k^2(\eta)}{\tilde{v}_k^2(\eta_0)}\, .
\end{equation}
Perturbations deep inside the Hubble radius have $\tilde{v}_k=1$.
If this value was constant throughout the entire inflationary evolution,
the solution to eq.~(\ref{Rckeq2}) would have a familar form 
corresponding to standard single-field inflation:
\begin{equation}
\label{eq:ord}
    \R_0 = \frac{C}{a}e^{-\mathrm{i}k\eta}\left( 1-\frac{\mathrm{i}}{k\eta}\right) \, .
\end{equation}
If this solution was true throughout the entire inflationary dynamics, at late times, $\eta\to 0^-$ we would have 
\begin{align}
    |\R_0|^2 \sim |C|^2 \cdot \frac{\kappa^2H^2}{k^2}
\end{align}
However, with $\tilde{v}_k^2\propto a^p$, the true solution is (\ref{eq:Rsol}), whose late-time limit for $p<2$ leads to:
\begin{align}
    |\R|^2 \sim |D_1|^2 \left( \frac{\Gamma\left(\frac{3-p}{2-p}\right)}{\pi}\right)^2 (2-p)^\frac{6-2p}{2-p} \, .
\end{align}
Using the Wronskian condition (\ref{eq:wr3}) with $W(\eta_0)$
calculated with the solution (\ref{eq:ord}), valid in the sub-Hubble limit, we obtain:
\begin{align}
    |C|^2 = \frac{(2-p)\kappa^{1-p}H}{\pi V_0^2 k} |D_1|^2\, .
\end{align}
Hence the enhancement factor for power spectrum of the curvature perturbations $\mathcal{P}$ (in comparison to the power spectrum for
a slow-roll single-field model $\mathcal{P}_\mathrm{sf}$) reads:
\begin{align}
    \frac{\mathcal{P}}{\mathcal{P}_\mathrm{sf}} = \frac{|\R|^2}{|\R_0|^2} = \left( \frac{k}{H}\right)^{\frac{p}{p-2}} V_0^{\frac{2}{p-2}} \frac{\Gamma^2\left(\frac{3-p}{2-p}\right)}{\pi}(2-p)^{\frac{4-p}{2-p}} \, .
\label{eq:enhf}
\end{align}
Eq.~(\ref{eq:enhf}) reproduces several well-known results. For $p=0$ and $V_0^2=\mathrm{const}$, corresponding to the first of the two limits discussed in Section \ref{sec:strongly}, we obtain:
\begin{align}
    \frac{\mathcal{P}}{\mathcal{P}_\mathrm{sf}} = \frac{1}{V_0} \, .
    \label{eq:enhf1}
\end{align}
For $p=-2$ and $V_0^2=\frac{k^2}{4\eta_\perp^2H^2}$, which correspond to the second limit in Section \ref{sec:strongly}, we have
\begin{align}
    \frac{\mathcal{P}}{\mathcal{P}_\mathrm{sf}} = \frac{8\sqrt{2}\,\left(\Gamma\left(\frac{5}{4}\right)\right)^2}{\pi}\,\eta_\perp^{1/2} \sim 2.96\,\eta_\perp^{1/2} \, . 
    \label{eq:enhf2}
\end{align}
This formula agrees very well with numerical results presented in \cite{Cremonini:2010ua}.

\begin{figure}
    \includegraphics[width=.44\textwidth]{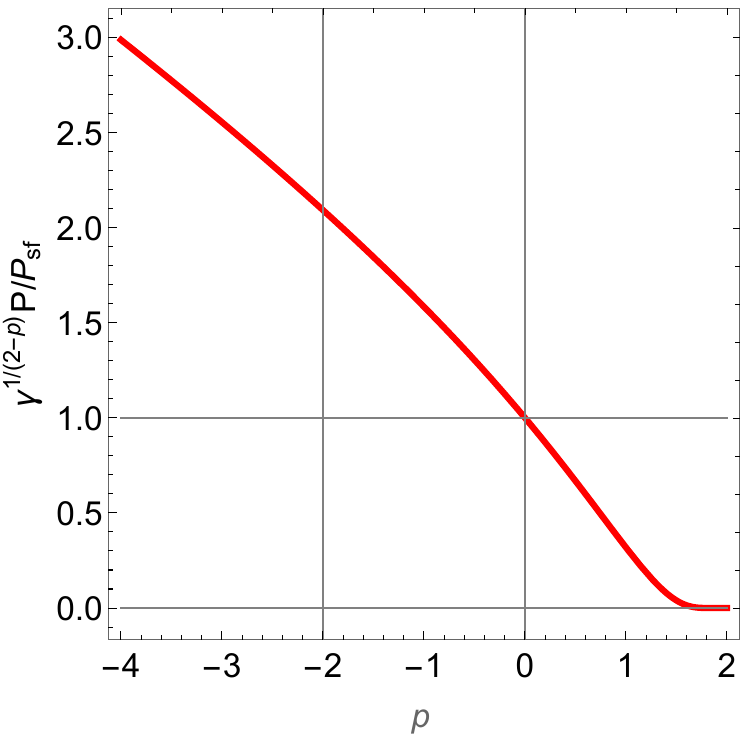}
    \caption{Enhancement of the power spectrum of curvature
    perturbations predicted by eq.~(\ref{eq:enhf}).
    }
\label{fig:enhf}
\end{figure}

Both results (\ref{eq:enhf1}) and (\ref{eq:enhf2}) correspond
to a scale-invariant power spectrum. Generally, if we parametrize $V_0^2 = \gamma (k/H)^q$, where $\gamma$
is a $k$-independent coefficient, 
the scalar spectral index is 
\begin{align}
n_s=1-\frac{p+q}{2-p} \, .    
\end{align}
Assuming
a scale-invariant power spectrum, i.e.~$p+q=0$, we show the predictions of the formula (\ref{eq:enhf}) in Figure~\ref{fig:enhf}. 

The calculations for the single-field case presented in this Section can
be easily generalized to a multi-field system. In Appendix \ref{sec:mf},
we present an appropriate derivation, followed in Appendix \ref{app:bdini} 
by a prescription
for matching the perturbations in the sub- and super-Hubble regime.

\section{Numerical examples}
\label{sec:num}

In Section \ref{sec:eff}, we have put forth a number a hypotheses. 
We argued that slow-roll fast-turn two-field inflationary models
can be effectively described by a single-field theory with
a time and $k$-dependent sound speed. We also proposed
which combination of modes serves as an effective degree of freedom
in the single-field theory. In this Section, we would like to
corroborate those findings by presenting results of numerical calculations.

We study the evolution of the perturbations in the model described by the Lagrangian:
\begin{equation}
\label{eq:Lnoncal}
\mathcal{L} = \frac{e^{-2\phi_2/M}}{2}(\partial\phi_1)^2-V_\mathrm{inf}(\phi_1) + \frac{1}{2}(\partial\phi_2)^2-\frac{1}{2}m_2^2\phi_1^2 \, .
\end{equation}
In this model, the interactions stemming from
the non-canonical kinetic term can compensate the potential force
acting on the field~$\phi_2$. As a consequence, there
may exist an inflationary trajectory, for which $\phi_1$ rolls
slowly and $\phi_2$ stays constant. Models of this type have been analyzed by many authors and it was found 
that for certain values of the parameters one can describe the curvature perturbations with a single-field effective theory, either one with an effective sound speed smaller than one or one with modified dispersion relations. 

Here we consider the approximation of quasi-de Sitter space, {\em i.e.}, following \cite{Cremonini:2010ua,Cremonini:2010sv}, we assume that the Hubble parameter is practically constant and
that the
field $\phi_1$ moves negligibly during inflation,
so all the quantities defined in terms of the homogeneous background
are also practically constant. In this approximation, 
equations of motion resulting from ({\ref{eq:Lnoncal}}) assume
the form (\ref{eq:diff2}) with (\ref{eq:ldef}) and (\ref{eq:mdef}),
where $\eta_\perp =\frac{\dot{\phi}_1}{MH}$ can be much larger than~1.
This approximation allows us to capture characteristic features 
of the evolution of the effective sound speed in various models
with high accuracy (which is particularly important 
for $|\eta_\perp|\gg1$), 
disentangling the effects of the changes 
in the sound speed from other time dependencies, e.g.\ those
originating from time-dependent background.
Of course, the MESS approach is completely general and does not require
the simplifications discussed here, but our goal is to discuss it
in the context of multi-field examples already worked out in the literature.

For numerical calculations, we use initial conditions (\ref{eq:solsubH})
and (\ref{eq:solsubHadd}) with $\theta_0=0$, integrating the equations
of motion (\ref{eq:diff2}) with (\ref{eq:ldef}) and (\ref{eq:mdef}) twice:
to cover both intitial conditions. In order to isolate the adiabatic mode
that dominates after Hubble radius crossing, we preform the following
unitary transformations of the two results corresponding to initial
conditions. If the first initial condition leads to $u_\sigma^{(1)}=z_1$
and the second initial 
condition leads to $u_\sigma^{(2)}=z_2$, 
we consider combinations of the two solutions, corresponding
to rotated vectors in (\ref{eq:solsubH}): 
\begin{equation}
    \left( \begin{array}{c}\tilde{u}_\sigma^{(1)} \\ \tilde{u}_\sigma^{(2)} \end{array} \right) =  \frac{1}{\sqrt{|z_1|^2+|z_2|^2}} \left( \begin{array}{cc}
         z_1^\ast & z_2^\ast \\
         -z_2 & z_1 
    \end{array}\right) \left( \begin{array}{c}u_\sigma^{(1)} \\ u_\sigma^{(2)} \end{array} \right) \, .
\end{equation}
At the end of numerical evolution, we have $\tilde{u}_\sigma^{(2)}\to 0$, 
and therefore we identify the {\em freezing} mode with $\tilde{u}_\sigma^{(1)}$ and
the {\em decaying} mode with $\tilde{u}_\sigma^{(2)}$.
According to our discussion in Appendix~\ref{sec:mf}, 
with freeze-in at sub-Hubble scales the freezing mode should
correspond to $z_2=-\mathrm{i}z_1$ and we confirm this in
our numerical examples.

We represent perturbations as instantaneous power spectra
and normalize them to the corresponding instantaneous power spectra
of curvature perturtbations in single-field models, as described
in detail in \cite{Lalak:2007vi}. We use color coding for different
components and different initial conditions described in Table~\ref{tab1}.

\begin{table}
    \centering
    \begin{tabular}{|p{0.15\textwidth}|p{0.5\textwidth}|p{0.2\textwidth}|}
         \hline
         \multicolumn{3}{|c|}{multi-field model} \\
         \hline
         perturbation & mode (defined by the behavior of the adiabatic mode) & color coding \\
         \hline
         curvature & freezing & thick, black, solid \\
         curvature & decaying & thin, black, dashed \\
         entropy & freezing & thin, red, dashed \\
         entropy & decaying & thin, red, solid \\
         \hline
         \multicolumn{3}{|c|}{single-field effective model} \\
         \hline
         curvature & $\tilde{v}_k$ given by eq.~(\ref{ck}) evaluated for the solution
    of the equations of motion corresponding to the freezing adiabatic
    mode & thick, green, dashed \\
         \hline
        curvature & $\tilde{v}_k$ given by eq.~(\ref{ck}) evaluated for the solution
    of the equations of motion corresponding to the decaying adiabatic
    mode & thick, yellow, dashed (only Fig.~\ref{fig:n2-4}) \\
         \hline
    \end{tabular}
    \caption{Color coding of the perturbations in Figures \ref{fig:n2-1}-\ref{fig:n2-4} }. 
    \label{tab1}
\end{table}

\subsection{Single-field effective theories with reduced sound speed}
\label{sec:ss-1}

\begin{figure}
    \includegraphics[width=.44\textwidth]{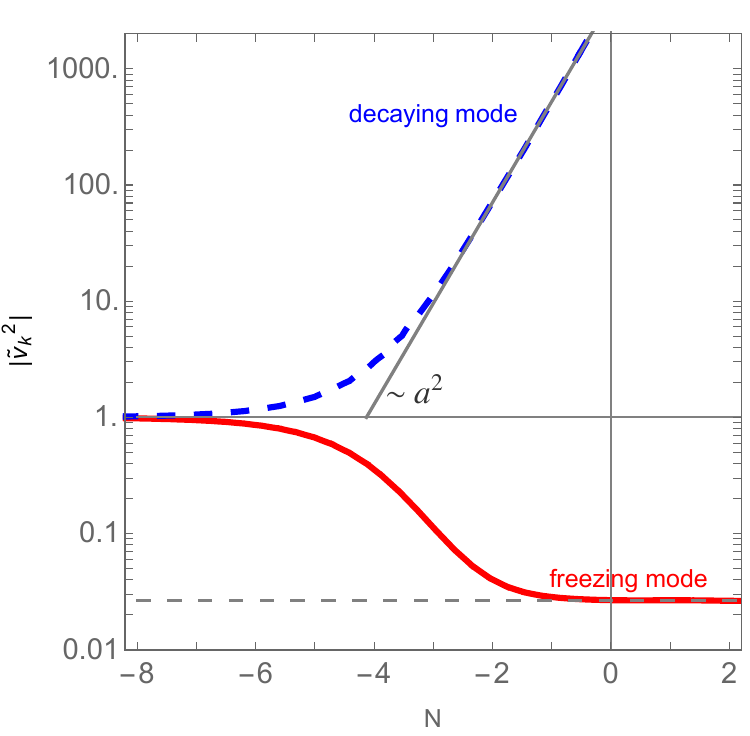}
    \hspace{0.1\textwidth}
    \includegraphics[width=.44\textwidth]{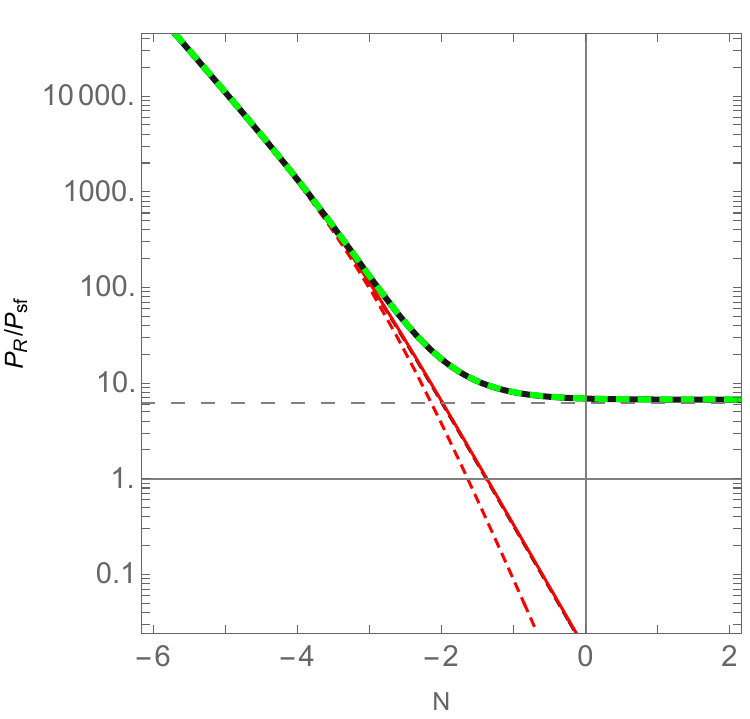}
    \caption{Numerical calculations in single-field effective 
    theories with constant reduced sound speed; model 
    described in Section~\ref{sec:ss-1}.
    Left panel: 
    evolution of the sound speed given by eq.~(\ref{ck}) for initial
    conditions leading to a freezing adiabatic mode (red solid line)
    and for initial conditions leading to a decaying adiabatic mode
    (blue dashed lines); thin dashed line corresponds to the value 
    (\ref{eq:vsgel2a}). Right panel: evolution of the instantaneous
    power spectra in the full theory and in the effective theory;
    color coding described in Table~\ref{tab1};
    thin dashed line
    corresponds to the asymptotic value (\ref{eq:enhf1}). $N=0$
    corresponds to the Hubble radius crossing
    }
\label{fig:n2-1}
\end{figure}

For the first numerical example, we assume $\eta_\perp=30$
and $\nu=10^2$, which leads to the effective sound speed
$\tilde{v}_k^2=0.0265\approx 1/37.7$. 
Evolution of the effective sound speed calculated from~(\ref{ck}) and evolution of adiabatic perturbations is shown in Figure~\ref{fig:n2-1}.
We find exquisite consistency at all scales 
between the predictions of the full
two-field model and the effective single-field theory
with a MESS sound speed.

\subsection{Single-field effective theories with modified dispersion relations}
\label{sec:ss-2}

\begin{figure}
    \includegraphics[width=.44\textwidth]{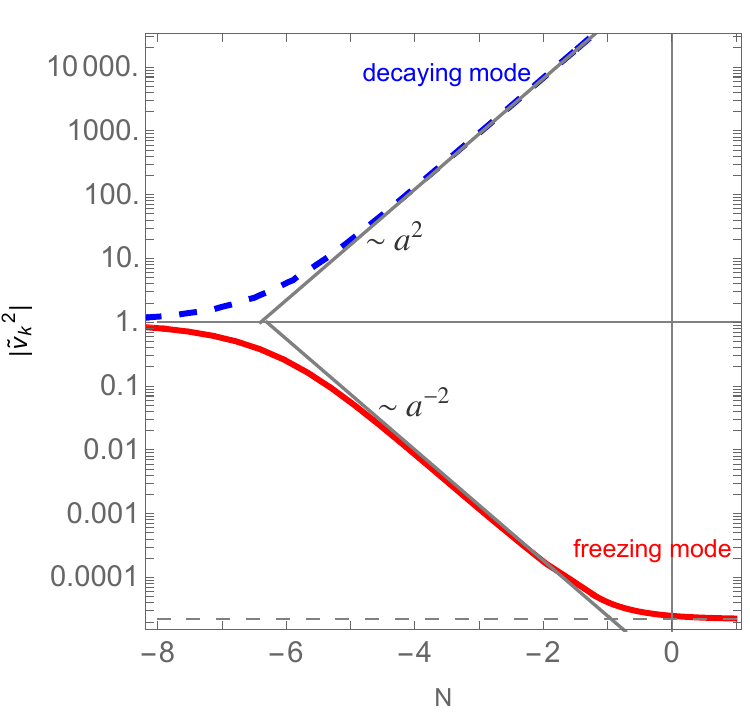}
    \hspace{0.1\textwidth}
    \includegraphics[width=.44\textwidth]{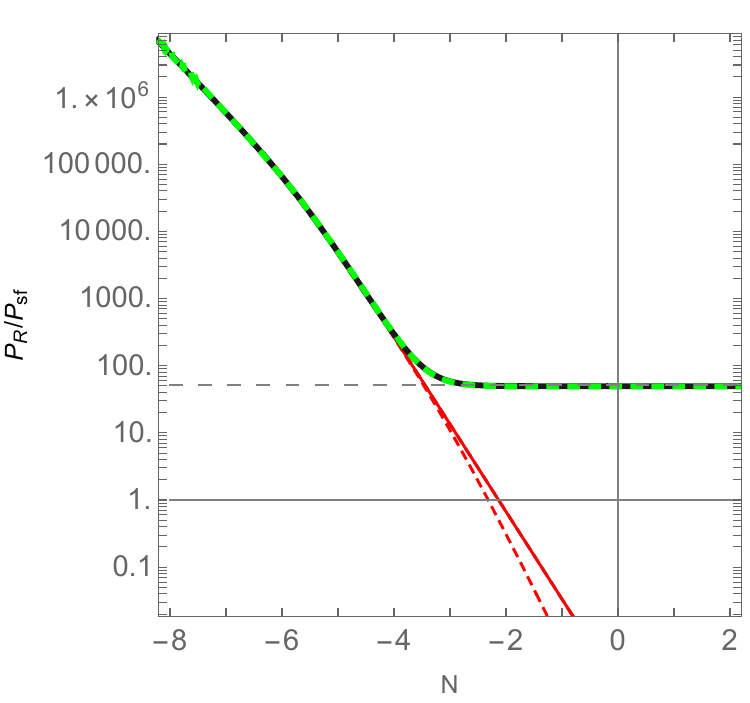}
    \caption{Numerical calculations in single-field effective 
    theories with modified dispersion relations; model 
    described in Section~\ref{sec:ss-2}.
    Left panel: 
    evolution of the sound speed given by eq.~(\ref{ck}) for initial
    conditions leading to a freezing adiabatic mode (red solid line)
    and for initial conditions leading to a decaying adiabatic mode
    (blue dashed lines); of thin dashed lines, one corresponds to the value 
    (\ref{eq:vsgel2a}) and the other shows that the sound speed
    decreases as $\sim a^{-2}$. Right panel: evolution of the instantaneous
    power spectra in the full theory and in the effective theory;
    color coding described in Table~\ref{tab1};
    thin dashed line
    corresponds to the asymptotic value (\ref{eq:enhf2}). $N=0$
    corresponds to the Hubble radius crossing
    }
\label{fig:n2-2}
\end{figure}

For the second numerical example, we assume $\eta_\perp=300$
and $\nu=10$.
This model is not described by an
effective single-field theory with a constant,
reduced sound speed, but rather by by an
effective single-field theory with modified dispersion
relations.
Evolution of the effective sound speed calculated from~(\ref{ck}) and evolution of adiabatic perturbations is shown in Figure~\ref{fig:n2-2}.
We again find exquisite consistency at all scales 
between the predictions of the full
two-field model and the effective single-field theory
with a MESS sound speed.

\subsection{Hyperinflation}
\label{sec:ss-3}

If the Lagrangian mass term for the entropy perturbations is small
compared to other scales, the mass of these perturbations is dominated
by the `geometrical' $-2H^2\eta_\perp^2$ term,
which in our example is related to the negative
curvature of the field space. 
Such a negative mass term leads to instability and to
a very strong enhancement of the amplitude of the perturbations.
This phenomenon was first described in \cite{Cremonini:2010ua}, which dubbed it
{\em transient tachyonic instability around the Hubble radius}, and
after a decade it was rediscovered in \cite{Brown:2017osf},
which called it {\em hyperinflation}, and further analyzed
in \cite{Mizuno:2017idt}. In a slightly different context,
{\em sidetracked inflation} models with a negative effective 
sound speed were discussed in \cite{Garcia-Saenz:2018ifx,Fumagalli:2019noh}. 
In all works mentioned above, inflation was realized on a steep potential in a hyperbolic field space.

\begin{figure}
    \includegraphics[width=.44\textwidth]{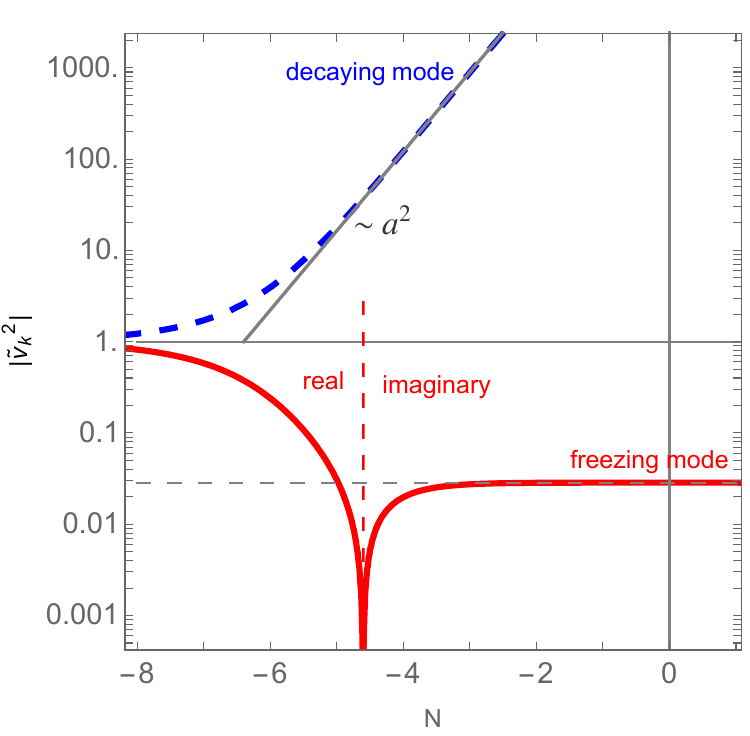}
    \hspace{0.1\textwidth}
    \includegraphics[width=.44\textwidth]{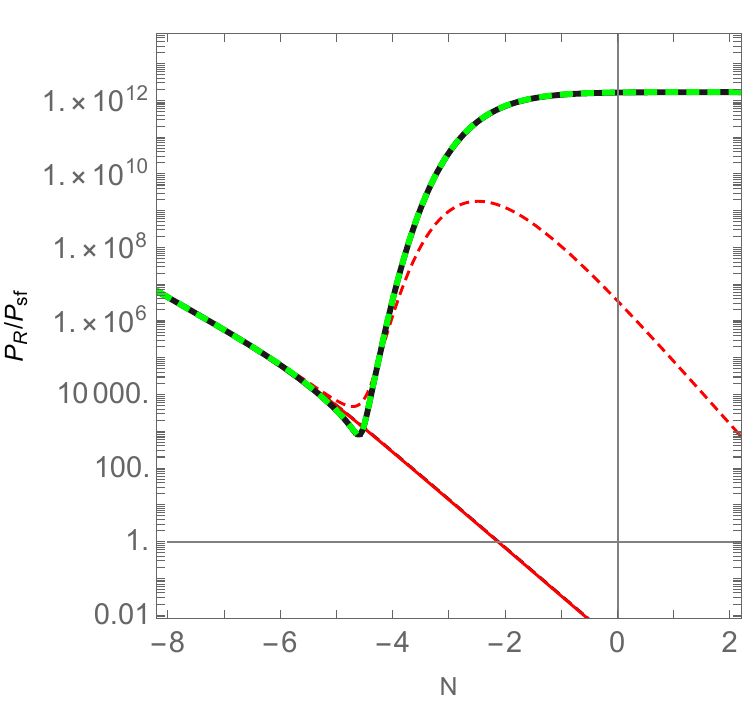}
    \caption{Numerical calculations in single-field effective 
    theories for hyperinflation; model 
    described in Section~\ref{sec:ss-3}.
    Left panel: 
    evolution of the sound speed given by eq.~(\ref{ck}) for initial
    conditions leading to a freezing adiabatic mode (red solid line)
    and for initial conditions leading to a decaying adiabatic mode
    (blue dashed lines); change of sign of the sound speed squared,
    i.e.~transition from real to imaginary sound speed,
    is indicated;
    thin dashed line corresponds to the value 
    (\ref{eq:vsgel2a}). Right panel: evolution of the instantaneous
    power spectra in the full theory and in the effective theory;
    color coding described in Table~\ref{tab1};
    $N=0$
    corresponds to the Hubble radius crossing
    }
\label{fig:n2-3}
\end{figure}

It is interesting to note that hyperinflation
can also be described in our effective single-field
approach, albeit with a sound speed $\tilde{v}_k^2$
which changes sign during evolution. We demonstrate this numerically
by an example with $\eta_\perp=300$ and $\nu=-10^4$. 
Evolution of the effective sound speed calculated from~(\ref{ck}) and evolution of adiabatic perturbations is shown in Figure~\ref{fig:n2-3}.
We find exquisite consistency at all scales 
between the predictions of the full
two-field model and the effective single-field theory
with a MESS sound speed.

In \cite{Cremonini:2010ua}, hyperinflation was described as an intrinsically
two-field phenomenon. However, \cite{Brown:2017osf} hinted at a curious
property, determined numerically, 
that the freezing adiabatic mode is obtained 
from a single, well-defined initial mode. Here we confirm
this observation and show that the evolution of that
mode can be understood in effective theory with
a time-dependent sound speed that starts at a canonical value
of~1 and then goes imaginary.

\subsection{Single-field description for models with light entropy modes that cannot be integrated out}
\label{sec:ss-4}

In Section~\ref{sec:eff}, 
we showed how the MESS approach allows to formulate a single-field 
description for models which were previously studied in the literature by integrating out entropy modes.
Here we will consider the case in which the approach based on integrating out  entropy modes cannot be applied, {\em i.e.} the first two terms in eq.(\ref{Qsigma}) cannot be neglected, and no simple algebraic relation between $\zeta$ and $Q_{\sigma}$ holds at all times. In these cases the MESS approach can still be used to compute the  effective sound speed of each independent quantum degree of freedom of the system. While from a field theoretic point of view  the fact that there are two light degrees of freedom would be interpreted as the non existence of a single-field description, the effective sound of the appropriately rotated modes allows to compute the final value of the full curvature spectrum by studying the evolution of a single degree of freedom, providing a single-field description.

We consider a model with light entropy perturbations,
$\nu=0$ and moderate kinetic coupling between perturbations, 
$\eta_\perp=0.3$. Such models were proposed in \cite{Cremonini:2010sv} to explain
in an alternative way the red tilt of the power spectrum
of adiabatic perturbations; later they were rediscovered
and analyzed anew in an improved way, invoking symmetries
of the theory \cite{Achucarro:2016fby}. Our particular model has entropy
perturbations slowly decaying, so the sourcing of the adiabatic
perturbations eventually becomes ineffective;
had we chosen $\nu=-2\eta_\perp^2$,
the amplitude of entropy perturbations would remain nearly constant
and the sourcing could last much longer. 

In these models, adiabatic perturbations are sourced by entropy
perturbations on super-Hubble scales, which corresponds to the situation
described in Section \ref{sec:sourcing}, with the sound speed
diverging to infinity. A closer inspection shows \cite{Cremonini:2010sv} that 
the amplitude of the adiabatic perturbations grows as $\sim\eta_\perp N$
on super-Hubble scales, hence the sound speed increases as
$\sim a^{2-\eta_\perp}$, according to eq.~(\ref{vk2}). 
Similarly, the sound speed for of the decaying mode increases as
%
%
$a^{2+\eta_\perp}$. This is consistent with our findigs in Section \ref{sec:sol} that $\tilde{v}_k\sim a^2$ marks a divide between freezing and
decaying solutions.

\begin{figure}
    \includegraphics[width=.44\textwidth]{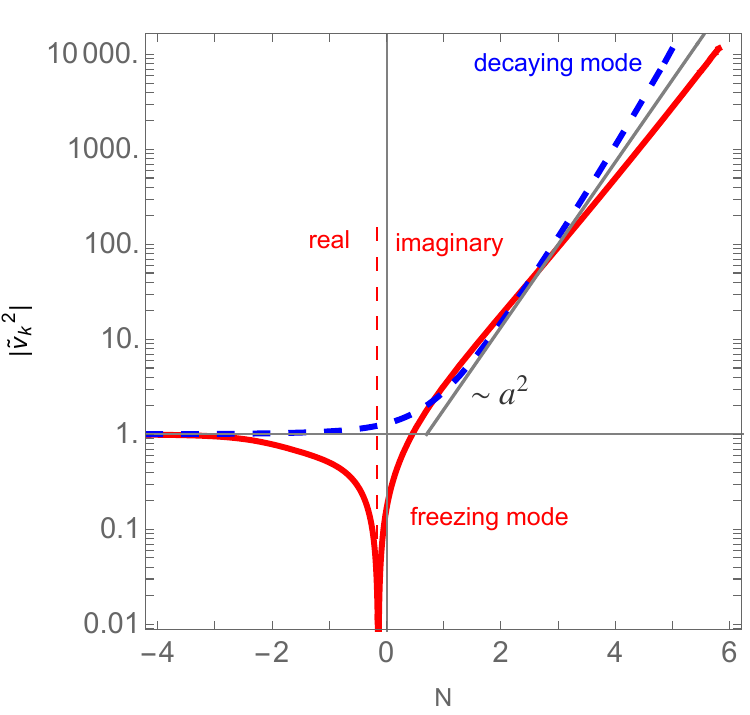}
    \hspace{0.1\textwidth}
    \includegraphics[width=.44\textwidth]{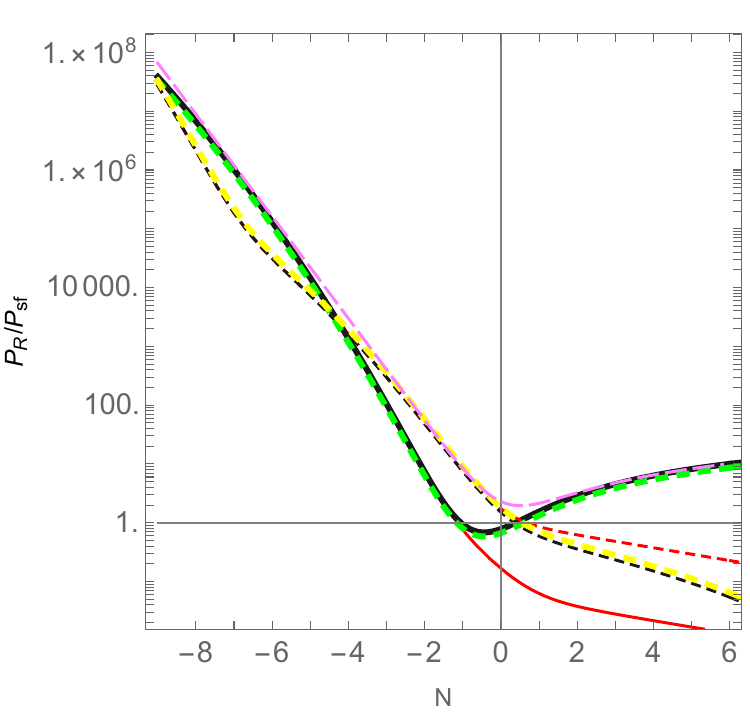}
    \caption{Numerical calculations in single-field effective 
    theories for light entropy perturbations; model 
    described in Section~\ref{sec:ss-4}.
    Left panel: 
    evolution of the sound speed given by eq.~(\ref{ck}) for initial
    conditions leading to a freezing adiabatic mode (red solid line)
    and for initial conditions leading to a decaying adiabatic mode
    (blue dashed lines); change of sign of the sound speed squared,
    i.e.~transition from real to imaginary sound speed,
    is indicated;
    thin dashed line corresponds to the value 
    (\ref{eq:vsgel2a}). Right panel: evolution of the instantaneous
    power spectra in the full theory and in the effective theory;
    color coding described in Table~\ref{tab1};
    $N=0$
    corresponds to the Hubble radius crossing; the pink line
    corresponds to the total curvature power spectrum.
    }
\label{fig:n2-4}
\end{figure}

In Figure~\ref{fig:n2-4}, we show that, similarly to the case
of hyperinflation, the sound speed $\tilde{v}_k^2$
changes sign during evolution.  
We also show the evolution of adiabatic and entropy perturbations.

The evolution of the freezing and decaying modes of the
adiabatic perturbations is compared to the evolution of 
a single-field effective description with an effective sound
speed given by (\ref{ck}) with an appropriate set of initial
conditions. 
We find a good agreement betwen the predictions of the full theory
and {\em two} single-field effective theories 
with different effective sound speeds. 
Depending on the phase of the evolution,
either the freezing or the decaying mode dominates the
instantaneous power spectrum and the late-time domination
of the freezing mode starts only after Hubble radius crossing.
This shows that the model {\em cannot} be approximated by
an effective single-field theory at all times -- we need to combine {\em two}
single-field theories with {\em two} effective, independent sound
speeds to obtain correct predictions for the curvature
perturbations at all times, but the freezing mode is sufficient to compute the final value.
Our numerical analyses also point to the fact that this
conclusion holds true for all models described in Section~\ref{sec:sourcing},
i.e.~models with sourcing of the adiabatic perturbations on super-Hubble
scales. 

Since those models can be studied also by integrating out entropy modes, the fact that a single EFT valid at any time does not exist, would also be a limitation of the EFT obtained using that method, and is an intrinsic property of these systems, independent of the method adopted to study them.  

\section{Discussion}
\label{sec:diss}

In the context of cosmological perturbations, the existence of
    a single-field effective theory requires that the degree of freedom
    corresponding to the freezing mode, accounting for the entire
    amplitude of
    adiabatic perturbations at the end of inflation, evolves independently
    of all other perturbations. Those perturbations can be dynamical, but
    as their masses are larger than the Hubble parameter, 
    their amplitudes decrease as power law functions of the scale
    factor. Hence the notion of the effective theory in cosmology is
    different from the one used in particle physics, where
    decoupling normally means that other degrees of freedom are too heavy
    to be excited.
    
At face value, our effective description of single-field
    inflation resembles the quadratic part of the action
    for adiabatic perturbations derived in \cite{Cheung:2007st}. 
    However, we would like to point out that the sound speed in that
    reference is a function of time only. Using a very simple model
    with a large and constant turning rate, analyzed previously
    in \cite{Cremonini:2010ua,Baumann:2011su}, we have shown
    that the evolution of the adiabatic perturbations is correctly
    accounted for by a sound speed that is both time- and 
    momentum-dependent. Hence our approach generalizes the {\em effective
    theory of inflation} of \cite{Cheung:2007st} in a non-trivial way, including the effects of entropy.
    

A truly effective single-field theory has only one relevant
degree of freedom that fully accounts for both the power spectrum
of the adiabatic perturbations and for higher-order correlation functions
of adiabatic perturbations.
Although such a mode has both the adiabatic and the entropic components,
a known effective sound speed (\ref{ck}) provides an algebraic
relation between these two components, so the entropic component
is no longer an independent quantity. Such an effective description
requires just one effective sound speed, because 
other degrees of freedom are assumed to have decayed before the
Hubble radius crossing and thus do not contribute to correlation
functions of adiabatic perturbations. In this sense, the models
analyzed in Sections \ref{sec:ss-1}-\ref{sec:ss-3} have a single-field
effective theory, while the model described in Section \ref{sec:ss-4} 
does not. In this latter case, there is a non-negligible independent
degree of freedom that significantly 
contributes to the amplitude of adiabatic perturbations around
the Hubble crossing. 
We can therefore conclude that a momentum-dependent effective sound
speed parametrizes single-field effective theories of inflation
and provides an effective description of the adiabatic perturbations
when such a theory cannot be formulated. 

There is also an alternative, more general view of the models discussed in 
Section \ref{sec:num}, which, however, involves more input and is thus less
predictive. 
Since the perturbed energy-stress tensor 
enters Einstein equations and does not rely
on a particular model of multi-field inflation, 
the evolution of the adiabatic component of each degree of freedom
is described by eq.~(\ref{Rckeq2}) with an appropriate sound speed.
We can define a number of different effective sound speeds
to account for the evolution of all degrees of freedom,
as we have done in Section~\ref{sec:ss-4}. This approach allows us to
describe also the evolution of adiabatic perturbations (without resorting
explicitly to the notion of entropy perturbations) in models which do not admit an effective single-field theory.

The effective field theory of inflation \cite{Cheung:2007st} 
is based on the assumption that only one scalar degree of freedom 
is present, and is formulated in the uniform field gauge, 
also called unitary gauge, in which an action invariant under 
time-dependent space diffeomerphism  can be  written without any 
matter perturbation terms. The unitary gauge does not coincide with 
the comoving slices gauge in multi-field systems \cite{Romano:2018frb}, 
so in general the effective theory of inflation cannot be applied 
to multi-field systems in which there is no gauge in which 
the matter perturbations can be completely set to zero (in other words,
 entropy perturbations) cannot be neglected. 
 Nevertheless, there can also be effective entropy 
 in the comoving slicea gauge in modified gravity theories 
 with a single scalar degree of freedom, e.g.\ in such as KGB models
\cite{Vallejo-Pena:2019hgv},  
which  can be described by the effective theory of inflation. 
These modified gravity theories give rise to a modification of the 
dispersion relation, related to extrinsic curvature terms 
of the effective action \cite{Cheung:2007st,Baumann:2011su} 
and leading to a momentum-effective sound speed,
consistent with the MESS approach, once the gauge transformation 
from the unitary to the comoving slices gauge is performed \cite{Vallejo-Pena:2019hgv}. 
In contrast, effective theory of inflation cannot be applied to multi-fields systems where there is no gauge in which the action can only be written in term of geometrical quantitites  This is confirmed, {\em e.g.} 
by the modified dispersion relation obtained in eq.~(B.6) in 
\cite{Baumann:2011su}, which has a different momentum dependency from the 
one which arises from extrinsic curvature terms in the effective theory of 
inflation, as shown in eq.~(3.22) in \cite{Baumann:2011su}, associated to 
intrinsic entropy in single field modified gravity theories.

In summary,  the advantages of the MESS are
that it relates the effective sound speed to the energy-stress tensor 
in a model-independent way. It also does not require integrating out  e
ntropy modes and it is not based on any further approximation, 
such as the decoupling limit often assumed 
in the effective theory of inflation.
Thus it gives a general model-independent description of adiabatic 
perturbations, valid at any energy scale. It also makes explicit the relation between the entropy of the mulfi-field theory 
and the momentum dependent effective sound speed of the corresponding single field effective theory, and that it can be computed directly from the solutions of the matter perturbations equations without the need of computing an effective action.


The definition of MESS is completely general, and can be applied to any multifields model, including models with sharp turns of the classical field trajectory.
It can also be applied to  modified gravity theories \cite{Vallejo-Pena:2019hgv}, and  more complex systems involving gauge fields, such as axion inflation, as long as the comoving gauge of the total effective energy-momentum tensor is properly computed. The ungauged tensor can always be computed analytically, while the comoving gauge condition can be added to the field equations to be solved numerically, in case it cannot be used to simply them analytically.

As long as numerical calculations can be carried out with sufficient accuracy, the method can be applied without any restrictions to any multi-field model, with no restriction on the classical field trajectory. The computation of the MESS involves in some cases the cancellation between very small numbers, which requires the use of a sufficiently high numerical accuracy to avoid instabilities, but for models where entropy modes cannot be integrated out, this is the only approach which can be adopted to obtain a single-field description capable of predicting the time dependence of the adiabatic perturbations of the full multi-field theory.


\section{Conclusions}
\label{sec:con}

In this work, we presented a formulation of a single-field
effective description of inflation, making use of a recently advocated
approach based on the momentum-dependent effective sound speed (MESS) \cite{Romano:2018frb}. We have shown that this formulation
includes a number of multi-field models that were considered
in the literature in the last decade. We have identified the effective
degree of freedom and shown how its evolution can be treated
independently of other degrees of freedom, even at scales
at which the amplitudes the latter are not suppressed yet.
We have also applied the MESS approach to a models with light
entropy perturbations, which does not admit an effective
field theory obtained by integrating out entropy modes.
Hence we have demonstrated that the MESS approach, which generalizes
the notion of single-field effective theory of inflation,
is a powerful and useful scheme for studying a wide range
of inflationary models.

\section*{Acknowledgments}
A.E.R.~thanks Juan Garcia Bellido for helpful discussions. A.E.R.~is partially supported by the National Agency of Academic Exchange (NAWA)
through its Ulam program.
K.T~is partially suported by the National Science Centre (NCN)
SHENG grant UMO-2018/30/Q/ST9/00795.

\appendix
\section{MESS of multiple scalar fields}
\renewcommand{\theequation}{\thesection.\arabic{equation}}

\label{MESSApp}

The  energy-stress tensor for the system described by the action given in eq.(\ref{action}) is 
\begin{equation}
    T^{\mu}{}_{\nu}=G_{IJ}\left(\Phi^K\right) \partial^{\mu} \Phi^I \partial_{\nu} \Phi^J + \delta^{\mu}{}_{\nu}\left[-\frac12 G_{IJ}\left(\Phi^K\right) \partial_{\lambda} \Phi^I \partial^{\lambda} \Phi^J-V\left(\Phi^K\right) \right] \, .
\end{equation}
The scalar fields at linear order can be expanded as $\Phi^K(x^{\mu})=\phi^K(t) + \delta \phi^K(x^{\mu})$, where the background parts of the scalar fields satisfy the following equations of motion
\begin{equation}
    \ddot{\phi}^I+3H\dot{\phi}^I + \Gamma^I_{JK}\dot{\phi}^J\dot{\phi}^K + G^{IJ}\left(\phi^K\right)V_{,J}\left(\phi^J\right) = 0 \, , \label{bfe}
\end{equation}{}
where $\Gamma^I_{JK}$ are the Christoffel symbols corresponding to the fields space metric $G_{IJ}\left(\phi^K\right)$, and we denote the partial derivative respect to the field $\phi^J$ according to $V_{,J}\left(\phi^J\right)=\frac{\partial }{\partial\phi^J}V\left(\phi^J\right)$.
The background energy density and pressure are  
\begin{align}
    \rho &= \frac{1}{2} \dot{\sigma}^2 + V\left(\phi^K\right) \, , \\
    P &= \frac{1}{2} \dot{\sigma}^2 - V\left(\phi^K\right) \, ,
\end{align}
where $\dot{\sigma}^2=G_{IJ}\left(\phi^K\right)\dot{\phi}^I\dot{\phi}^J$.
The components of the perturbed energy-stress tensor of the two scalar fields system, without gauge fixing, are  
\begin{align}
\delta T^{0}{}_{0} &= -\frac{1}{2}G_{IJ}\left(\phi^K\right)\left(\dot{\phi}^I \dot{\delta\phi}^J + \dot{\phi}^J \dot{\delta\phi}^I \right) + \dot{\sigma}^2  A -  \delta \phi^k \left( \frac{1}{2}\dot{\phi}^I\dot{\phi}^J G_{IJ},_K\left(\phi^K\right) + V_{,K}\left(\phi^K\right) \right) \nonumber \, ,  \\
\delta T^{i}{}_{j} &= \delta^{i}_{j} \left[  \frac{1}{2}G_{IJ}\left(\phi^K\right)\left(\dot{\phi}^I \dot{\delta\phi}^J + \dot{\phi}^J \dot{\delta\phi}^I \right) - \dot{\sigma}^2  A +  \delta \phi^k \left( \frac{1}{2}\dot{\phi}^I\dot{\phi}^J G_{IJ},_K\left(\phi^K\right) -V_{,K}\left(\phi^K\right) \right) \right] \, ,  \nonumber  \\
\delta T^{0}{}_{i} &=- \partial_i \left[ \frac{G_{IJ}\left(\phi^K\right)\dot{\phi}^I \delta \phi^J}{a} \right] \, . \label{pq}
\end{align}
Under an infinitesimal time translation $t \to t+\delta t$ the fields perturbations transform according to the gauge transformation
\begin{align}
\widetilde{\delta \phi}^K &= \delta \phi^K - \dot{\phi}^K \delta t  \, . \label{gtdeltaphi}
\end{align}
From these equations we can find the time translation $\delta t_c$ necessary to go to the comoving gauge, by imposing the comoving gauge condition $(\delta T^{0}{}_{i})_c=0 \rightarrow G_{IJ}\left(\phi^K\right)\dot{\phi}^I \widetilde{\delta \phi}^J = 0$,  obtaining 
\begin{align}
\delta t_c &= \frac{G_{IJ}\left(\phi^K\right)\dot{\phi}^I \delta \phi^J}{\dot{\sigma}^2} \, . \label{cgt}
\end{align}
We can now compute the gauge invariant comoving field perturbations according to 
\begin{equation}
    U^K = \delta \phi^K - \dot{\phi}^K \delta t_c = \delta \phi^K - \dot{\phi}^K \frac{G_{IJ}\left(\phi^K\right)\dot{\phi}^I \delta \phi^J}{\dot{\sigma}^2}  \, , \label{Udef}
\end{equation}{}
and the comoving pressure and energy density perturbations
\begin{align}
    \alpha &= \delta P_c =   \frac{1}{2}G_{IJ}\left(\phi^K\right)\left(\dot{\phi}^I \dot{U}^J + \dot{\phi}^J \dot{U}^I \right) - \dot{\sigma}^2  \gamma +  U^k \left( \frac{1}{2}\dot{\phi}^I\dot{\phi}^J G_{IJ},_K\left(\phi^K\right) -V_{,K}\left(\phi^K\right) \right)\, , \\
    \beta &= \delta \rho_c=   \frac{1}{2}G_{IJ}\left(\phi^K\right)\left(\dot{\phi}^I \dot{U}^J + \dot{\phi}^J \dot{U}^I \right) - \dot{\sigma}^2 \gamma +  U^k \left( \frac{1}{2}\dot{\phi}^I\dot{\phi}^J G_{IJ},_K\left(\phi^K\right) + V_{,K}\left(\phi^K\right) \right) \, .
\end{align}{}
After replacing eq.(\ref{Udef}) and eq.(\ref{bfe}) into these expressions we find 
\begin{align}
     U^k V_{,K}\left(\phi^K\right) &= \frac{1}{2}G_{IJ}\left(\phi^K\right)\left(\dot{\phi}^I \dot{U}^J + \dot{\phi}^J \dot{U}^I \right) +  U^k \frac{1}{2}\dot{\phi}^I\dot{\phi}^J G_{IJ},_K\left(\phi^K\right) = - \dot{\sigma}^2 \frac{\Theta}{4} \, , \\
    \alpha &= - \dot{\sigma}^2  \gamma = - \dot{\sigma}^2  \frac{\dot{\zeta}}{H} \, , \\
    \beta &= - \dot{\sigma}^2  \left(\gamma +\frac{\Theta }{2}\right) = - \dot{\sigma}^2  \left(\frac{\dot{\zeta}}{H} +\frac{\Theta}{2}\right)  \, , 
\end{align}
where we have used the perturbed Einstein's equation $\gamma=\dot{\zeta}/H$, and we have defined the function $\Theta$ according to 
\begin{align}
    \Theta &\equiv-\frac{4\dot{\phi}_1\dot{\phi}_2}{\dot{\sigma}^3} \sqrt{G} \left( \frac{\delta \phi_1}{\dot{\phi}_1}-\frac{\delta \phi_2}{\dot{\phi}_2}\right)  V_{,s} = \frac{4}{\dot{\sigma}^2}  Q_{,s} V_{,s} \, ,
\end{align}
where $G$ is the determinant of the fields space metric $G_{IJ}\left(\phi^K\right)$, i.e. $G\equiv \det{(G_{IJ})}$, $Q_{,s} \equiv Q_{,K}e^{K}_{s}$ $V_{,s} \equiv V_{,K} e^{K}_{s}$, and 
\begin{align}
    e^{K}_{s} &= \left( e^{1}_{s}, e^{2}_{s} \right) = \left( \frac{G_{21}\dot{\phi}_1+G_{22}\dot{\phi}_2}{\dot{\sigma}\sqrt{G}}, -\frac{G_{11}\dot{\phi}_1+G_{12}\dot{\phi}_2}{\dot{\sigma}\sqrt{G}} \right) \, .
\end{align}

Finally the MESS is given by 
\begin{align}
\tilde{v}_k^2(t) &= \left(1+\frac{H \Theta}{2 \dot{\R}}\right)^{-1} = \left(1+\frac{2 H V_{,s} Q_{,s}  }{\dot{\R} \dot{\sigma}^2}\right)^{-1} = \left(1-\frac{2 H^2 \eta_\perp Q_{,s}  }{\dot{\R} \dot{\sigma}}\right)^{-1} \, , \label{ss} 
\end{align}
where 
\begin{equation} 
\eta_\perp  \equiv -\frac{V_{,s} }{H \dot \sigma} \, .
\end{equation}

\section{Multi-field case}
\label{sec:mf}

The calculation given in Section~\ref{sec:lf1} can be easily generalized
to a system of $N$ coupled linear and homogeneous equations, which
can be written as:
\begin{equation}
\label{eq:diff2}
    \frac{\mathrm{d}^2\vec{\mathcal{U}}}{\mathrm{d}\eta^2} + \mathbb{L}(\eta)\frac{\mathrm{d}\vec{\mathcal{U}}}{\mathrm{d}\eta}+\mathbb{M}(\eta)\vec{\mathcal{U}} = 0 \,,
\end{equation}
where $\vec{\mathcal{U}}=(\mathcal{U}_1(\eta),\ldots,\mathcal{U}_N(\eta))$ and $\mathbb{L}(\eta)$,
$\mathbb{M}(\eta)$ are real-valued $N\times N$ matrices, which are
functions of the independent variable~$\eta$. 
It is easy to show that for $\mathbb{L}=0$ and $\mathbb{M}^T=\mathbb{M}$ 
the Wronskian
defined as:
\begin{equation}
\label{eq:wrmd}
    W(\eta) \equiv \vec{\mathcal{U}}^\dagger \frac{\vec{\mathcal{U}}}{\mathrm{d}\eta}-\left( \frac{\mathrm{d}\vec{\mathcal{U}}}{\mathrm{d}\eta}\right)^\dagger \vec{\mathcal{U}}
\end{equation}
does not depend on~$\eta$. 

The equations of motion for the two-field system of adiabatic and entropy perturbations (\ref{Qsigma})-(\ref{Qs}) can be transformed so that we can make use of this fact. We first redefine perturbations as $\vec{u}=(aQ_\sigma,aQ_s)$ and identify $\eta$ with conformal time. 
We obtain a system of equations of the form (\ref{eq:diff2}) with:
\begin{equation}
    \mathbb{L} = \left( \begin{array}{cc} 0 & \frac{2\eta_\perp}{\eta} \\ -\frac{2\eta_\perp}{\eta} & 0 \end{array} \right)
    \label{eq:ldef}
\end{equation}
and
\begin{equation}
    \mathbb{M} = \left( k^2-\frac{2}{\eta^2} \right)\,\mathbf{1}+ \left( \begin{array}{cc} 0 & -\frac{4\eta_\perp}{\eta^2} \\ -\frac{2\eta_\perp}{\eta^2} & \frac{\nu}{\eta^2} \end{array} \right) \, ,
    \label{eq:mdef}
\end{equation}
where $\nu=\frac{m_s^2}{H^2}-2\eta_\perp^2$ and we used de Sitter approximation again \footnote{This system was given e.g.~in \cite{Lalak:2007vi}
and \cite{Cremonini:2010ua}, but some later references \cite{Brown:2017osf,Mizuno:2017idt}
write these equations with
$\mathbb{M}^T$ instead of $\mathbb{M}$ without commenting on this
discrepancy.}. 
We then define
\begin{equation}
    \vec{\mathcal{U}} = \mathbb{R} \vec{u} 
\end{equation}
with
\begin{equation}
    \mathbb{R}(\eta) = \left( \begin{array}{cc} \cos\left( \eta_\perp\log\left(\frac{\eta}{\eta_0}\right) \right) & \sin\left( \eta_\perp\log\left(\frac{\eta}{\eta_0}\right) \right) \\ -\sin\left( \eta_\perp\log\left(\frac{\eta}{\eta_0}\right) \right)  & \cos\left( \eta_\perp\log\left(\frac{\eta}{\eta_0}\right) \right)  \end{array}\right) \,,
\end{equation}
where $\eta_0$ is an arbitrary constant. In terms of the new variable~$\vec{\mathcal{U}}$, the equation of motion~(\ref{eq:diff2})
reads:
\begin{equation}
    \frac{\mathrm{d}^2\vec{\mathcal{U}}}{\mathrm{d}\eta^2} + \left[ \left( k^2+\frac{\eta_\perp^2-2}{\eta^2}\right)\mathbf{1}+\frac{1}{\eta^2}\mathbb{R}\mathcal{M}\mathbb{R}^T \right]\vec{\mathcal{U}} = 0 \, ,
    \label{eq:diff2a}
\end{equation}
where
\begin{equation}
    \mathcal{M} = \left( \begin{array}{cc} 0 & -3\eta_\perp  \\ -3\eta_\perp &   \nu  \end{array} \right) \, ,
\end{equation}
The conserved Wronskian (\ref{eq:wrmd}) reads:
\begin{equation}
    W(\eta) = \vec{u}^\dagger \frac{\mathrm{d}\vec{u}}{\mathrm{d}\eta}-\left( \frac{\mathrm{d}\vec{u}}{\mathrm{d}\eta}\right)^\dagger \vec{u}+\frac{2\eta_\perp}{\eta} \vec{u}^\dagger \mathbb{E} \vec{u} \,,
    \label{eq:wru}
\end{equation}
where we denoted:
\begin{equation}
    \mathbb{E} = \left( \begin{array}{cc} 0 & 1 \\ -1 & 0 \end{array} \right)
\end{equation}
and made use of the fact that $\frac{\mathrm{d}\mathbb{R}}{\mathrm{d}\eta}=\frac{\eta_\perp}{\eta}\mathbb{R}\mathbb{E}$.

This form of the Liouville equation can be used to identify
the initial Bunch-Davis conditions in a coupled multi-field
system and to match those initial condition with the late-time
behavior of the perturbations. We comment on these issues below.

\section{Matching curvature and entropy perturbations in the sub- and super-Hubble regime}
\label{app:bdini}

Based on the results of Appendix~\ref{sec:mf},
we can comment on the choice of the Bunch-Davies
vacuum as an initial state for the adiabatic and entropy perturbations and on a simple way in which that initial state
can be matched with the asymptotic late-times solutions
of the equations of motion.
Deep inside the Hubble radius, i.e.~for $\eta\to-\infty$, eq.~(\ref{eq:diff2a}) becomes an equation of motion 
for a harmonic oscillator and 
it has two independent positive-frequency solutions:
\begin{equation}
    \vec{\mathcal{U}}^{(1)}(\eta) \sim \frac{e^{-\mathrm{i}k\eta}}{\sqrt{2k}}
    \vec{\mathcal{U}}^{(1)}_0
    \qquad\textrm{and}\qquad
    \vec{\mathcal{U}}^{(2)}(\eta) \sim \frac{e^{-\mathrm{i}k\eta}}{\sqrt{2k}}
    \vec{\mathcal{U}}^{(2)}_0 \, ,
    \label{eq:solsubH}
\end{equation}
where 
$\vec{\mathcal{U}}^{(1)}_0$ and $\vec{\mathcal{U}}^{(2)}_0$
are constant vectors satisfying
\begin{equation}
    \vec{\mathcal{U}}^{(I)\dagger}_0 \vec{\mathcal{U}}^{(J)}_0 = \delta_{IJ} \, .
\end{equation}
These vectors can be parametrized as:
\begin{equation}
    \vec{\mathcal{U}}^{(1)}_0 = \left( \begin{array}{c} \cos\theta_0 \\ \sin\theta_0 e^{i\phi_0} \end{array}\right)
       \qquad\textrm{and}\qquad
 \vec{\mathcal{U}}^{(2)}_0 = \left( \begin{array}{c} -\sin\theta_0 e^{-i\phi_0} \\ \cos\theta_0  \end{array}\right) \, .   
     \label{eq:solsubHadd}
\end{equation}
In terms of perturbations $\vec{u}$, the solution (\ref{eq:solsubH})
reads:
\begin{equation}
    \vec{u}^{(1)} \sim \frac{e^{-\mathrm{i}k\eta}}{\sqrt{2k}} \left( \begin{array}{c} \cos\theta_0 \cos\left( \eta_\perp\log\left(\frac{\eta}{\eta_0}\right) \right) -e^{i\phi_0}\sin\theta_0\sin\left( \eta_\perp\log\left(\frac{\eta}{\eta_0}\right) \right) \\ \cos\theta_0 \sin\left( \eta_\perp\log\left(\frac{\eta}{\eta_0}\right) \right) +e^{i\phi_0}\sin\theta_0\cos\left( \eta_\perp\log\left(\frac{\eta}{\eta_0}\right) \right) \end{array}\right)
    \label{eq:solsubH2}
\end{equation}
and
\begin{equation}
    \vec{u}^{(2)} \sim \frac{e^{-\mathrm{i}k\eta}}{\sqrt{2k}} \left( \begin{array}{c} -e^{-\mathrm{i}\phi_0}\sin\theta_0 \cos\left( \eta_\perp\log\left(\frac{\eta}{\eta_0}\right) \right) -\cos\theta_0\sin\left( \eta_\perp\log\left(\frac{\eta}{\eta_0}\right) \right) \\ -e^{-\mathrm{i}\phi_0}\sin\theta_0 \sin\left( \eta_\perp\log\left(\frac{\eta}{\eta_0}\right) \right) +\cos\theta_0\cos\left( \eta_\perp\log\left(\frac{\eta}{\eta_0}\right) \right) \end{array}\right)
    \label{eq:solsubH2a} \, .
\end{equation}
The modulus squared of the upper (adiabatic) component
in (\ref{eq:solsubH2}) reads:
\begin{equation}
    \left|u^{(1)}_\sigma\right|^2 = \frac{1}{4k} \left( 1 + \cos2\theta_0\cos\left( 2\eta_\perp\log\left(\frac{\eta}{\eta_0}\right) \right)-\cos\phi_0\sin2\theta_0\sin\left( 2\eta_\perp\log\left(\frac{\eta}{\eta_0}\right) \right) \right) \, .
\end{equation}
This expression is constant for $\theta_0=\pm\frac{\pi}{4}$ and $\phi_0=\frac{\pi}{2}$, which also corresponds to constant
$|u^{(1)}_s|^2$, $|u^{(2)}_\sigma|^2$ and $|u^{(2)}_s|^2$.
Our final results is, therefore:
\begin{equation}
     \vec{u}^{(1)} \sim \frac{e^{-\mathrm{i}k\eta+\mathrm{i}\eta_\perp\log\left(\frac{\eta}{\eta_0}\right)}}{2\sqrt{k}} \left( \begin{array}{c} 1 \\ -\mathrm{i}  \end{array}\right)
     \qquad\textrm{and}\qquad
     \vec{u}^{(2)} \sim \frac{e^{-\mathrm{i}k\eta-\mathrm{i}\eta_\perp\log\left(\frac{\eta}{\eta_0}\right)}}{2\sqrt{k}} \left( \begin{array}{c} -\mathrm{i} \\ 1  \end{array}\right) \, .
       \label{eq:solsubH3}
\end{equation}
Note that eq.~(\ref{eq:solsubH3}) exhibits some redundancy, which was not visible in the intermediate steps leading to that result. A change in arbitrary constant $\eta_0$ can be extracted as an unphysical phase
factor multiplying the solution.

The approximate solution~(\ref{eq:solsubH3}) is reliable as long as
the last term in eq.~(\ref{eq:diff2a}) is negligible. This is satified for
$(k\eta)^2>\mathrm{max}\{\nu,3\eta_\perp\}$.

It is also interesting to study the late-time behavior of
the system of equations (\ref{eq:diff2a}) with (\ref{eq:ldef})
and (\ref{eq:mdef}), following the treatment in \cite{Cremonini:2010ua}. 
In the limit $\eta\to0^-$, we can neglect
the $k$-dependent term and assume solutions of the form:
\begin{equation}
    \vec{u} = 
     \left( \frac{\eta}{\eta_0}\right)^P
    \left( \begin{array}{c}
         A_\sigma \\
         A_s
    \end{array} \right) \, ,
\end{equation}
where $\eta_0$ represents the value of the conformal time
at which the solution should be matched with the early-time
solution.
We obtain an algebraic equation:
\begin{equation}
    \left( \begin{array}{cc}
         P(P-1)-2 & 2\eta_\perp(P-2)  \\
         -2\eta_\perp(P+1) & P(P-1)-2+\nu
    \end{array}\right) 
    \left( \begin{array}{c}
         A_\sigma \\
         A_s
    \end{array} \right) = 0 \, .
    \label{eq:ltalg}
\end{equation}
Eq.~(\ref{eq:ltalg}) has four nontrivial solutions for $p$:
\begin{eqnarray}
\label{eq:solalg1}
P_1 &=& -1 \,, \qquad\textrm{with}\qquad \frac{A^{(1)}_s}{A^{(1)}_\sigma} = 0 \\
\label{eq:solalg2}
P_2 &=& 2 \,, \qquad\textrm{with}\qquad  \frac{A^{(2)}_s}{A^{(2)}_\sigma} = \frac{6\eta_\perp}{\nu} \\
P_{3,4} &=& \frac{1}{2}\mp \mathrm{i} \sqrt{\nu+4\eta_\perp^2-\frac{9}{4}} \,,
\qquad\textrm{with}\qquad 
\frac{A^{(3,4)}_s}{A^{(3,4)}_\sigma} = - \frac{\nu+4\eta_\perp^2}{\eta_\perp \left( 3\pm 2\mathrm{i}\sqrt{\nu+4\eta_\perp^2-\frac{9}{4}} \right)} \, .
\label{eq:solalg3}
\end{eqnarray}
The last two solutions (\ref{eq:solalg3}) correspond to the positive
and negative frequency solutions for
a massive mode, of mass squared $(\nu+4\eta_\perp^2)H^2$.
The first two solutions, eqs.~(\ref{eq:solalg1})-(\ref{eq:solalg2})
correspond to the growing and decaying part of a massless mode.
It is also clear
that the growing mode $\sim 1/\eta$ carries
only the adiabatic component, i.e.~in the considered model 
adiabatic perturbations can freeze in at some
scale, while all entropy perturbations decay at late times. 

The mode corresponding to the exponent $p_4$ corresponds 
to negative frequency. If the relative change of the sound speed
is not much larger than one, this mode is not excited during
the evolution of the perturbations. It is instructive to analyze
the relations between the sub-Hubble solutions (\ref{eq:solsubH3})
and the solutions~(\ref{eq:solalg1})-(\ref{eq:solalg3}). This is particularly
simple in the limit $\nu\to0$, which will correspond to
numerical examples to be discussed later. In this limit, we have:
\begin{equation}
    A_s^{(1)} = 0\,, \qquad A_\sigma^{(2)} \approx 0\,, \qquad A_s^{(3)} 
    \approx \mathrm{i} A_\sigma^{(3)} \, .
\end{equation}
Matching (\ref{eq:solsubH3}) with (\ref{eq:solalg1})-(\ref{eq:solalg3}),
we find that $\vec{u}^{(2)}$ corresponds to a massive mode with $p_3$, 
which decays on super-Hubble scales,
while $\vec{u}^{(1)}$ is a combination of a growing mode corresponding
to $p_1$ and the decaying massive mode corresponding to $p_2$, with
$A_\sigma^{(1)}\approx -\mathrm{i} A_s^{(2)}$. 

A general late-times solution of (\ref{eq:diff2}) can therefore
be written as:
\begin{equation}
    \vec{u} = \sum_{I=1}^4  \left( \frac{\eta}{\eta_0}\right)^{P_I}
    \left( \begin{array}{c}
         A^{(I)}_{\sigma} \\
         A^{(I)}_{s}
    \end{array} \right) \, ,
    \label{eq:sollt2}
\end{equation}
where for a given $I$ the coefficients $A^{(I)}_{\sigma}$ 
and $A^{(I)}_{s}$ satisfy the relations
in respective eqs.~(\ref{eq:solalg1})-(\ref{eq:solalg3}).
Plugging (\ref{eq:sollt2}) into the expression for the conserved Wronskian,
we find:
\begin{equation}
    W = -\frac{\mathrm{i}\left(\nu+4\eta_\perp^2\right)}{\eta_\perp\eta_0} \,\mathrm{Im}\left( A^{(1)}_{\sigma}A_{s}^{(2)\ast} \right) -\frac{\mathrm{i}\left(\nu+4\eta_\perp^2\right)\sqrt{\nu+4\eta_\perp^2-\frac{9}{4}}}{2\eta_\perp^2\eta_0} \left( \left|A^{(3)}_{\sigma}\right|^2 - \left|A^{(4)}_{\sigma}\right|^2 \right) \, .
\end{equation}
In the limit $\nu\to0$ considered above, this reduces to:
\begin{equation}
    W = - \frac{4\mathrm{i}\eta_\perp}{\eta_0} |A^{(1)}_{\sigma}|^2 \, .
    \label{eq:wru4}
\end{equation}
As the Wronskian (\ref{eq:wru4}) is conserved and equal $-\mathrm{i}$, 
we find that $|A_{\sigma,1}|^2=\eta_0/4\eta_\perp$, which leads 
to the following
prediction for the power spectrum of the adiabatic perturbations:
\begin{equation}
\frac{\mathcal{P}}{\mathcal{P}_\mathrm{sf}} = \frac{|k\eta_0|^3}{2\eta_\perp}
\, .
\end{equation}
Since $\eta_0$ corresponds to matching between the early- and late-time
solutions, and we argued that for $\nu\to0$ 
we have $\eta_0=-\sqrt{3\eta_\perp}/k$,
we obtain:
\begin{equation}
\frac{\mathcal{P}}{\mathcal{P}_\mathrm{sf}} = \frac{3\sqrt{3}}{2} \eta_\perp^{1/2} \, .
\label{eq:uga}
\end{equation}
We note that this equation has the same parametric form as
eq.~(\ref{eq:enhf2}) and the numerical prefactor $\sim2.6$ in 
eq.~(\ref{eq:uga}) is very close to that eq.~(\ref{eq:enhf2}).
This is a remarkable consistency, given our crude approach to
solving the equations of motion for the two-field system, relying
on matching between the early- and late-time asymptotic solutions.


\bibliographystyle{h-physrev4}
\bibliography{mybib}
\end{document}